\documentclass[twocolumn,pra,amsmath,amssymb,unsortedaddress,showpacs]{revtex4}

\usepackage{graphicx, amsmath}
\usepackage{verbatim}

\newcommand{\vect}[1]{\boldsymbol{#1}}
\newcommand{\unitvect}[1]{\boldsymbol{\hat{#1}}}
\newcommand{\Uminus}{$U_-$}
\newcommand{\Uplus}{$U_+$}
\newcommand{\ket}[1]{|#1\rangle}
\newcommand{\bra}[1]{\langle #1|}

\newcommand{\ToFrom}{\leftrightarrow}
\newcommand{\matrixel}[3]{\bra{#1}#2\ket{#3}}

\begin{document}

\title{Bose-Einstein condensates in RF-dressed adiabatic potentials}

\author{M. White, H. Gao, M. Pasienski, B. DeMarco}

\affiliation{Department of Physics, University of Illinois at
Urbana-Champaign, Urbana, IL 61801, USA}

\date{\today}


\begin{abstract}
Bose-Einstein condensates of $^{87}$Rb atoms are transferred into
radio-frequency (RF) induced adiabatic potentials and the
properties of the corresponding dressed states are explored.  We
report on measurements of the spin composition of dressed
condensates.  We also show that adiabatic potentials can be used
to trap atom gases in novel geometries, including suspending a
cigar-shaped cloud above a curved sheet of atoms.
\end{abstract}

\maketitle

\section{Introduction}

Adiabatic potentials, created using a combination of
radio-frequency (RF) and static magnetic fields, are promising
tools for confining ultra-cold atoms in novel geometries and for
use as beamsplitters in atom interferometry.
Examples of proposed trapping geometries include potentials with
two dimensional character \cite{Zobay01,Zobay04,Lorent04}, ring
shaped potentials \cite{Garraway05}, and tunnel arrays
\cite{Zimmermann06}. The use of adiabatic potentials as a
beamsplitter in an atomic interferometry application was recently
demonstrated in \cite{Schmiedmayer05}. A potential advantage of
this technique for exploring atomic superfluids in new geometries
over other methods (e.g., optical traps) is that the potential can
be very smooth \cite{endnote1}.

In this work, we investigate the properties of ultra-cold atoms
confined in adiabatic potentials generated through a combination
of RF magnetic fields, applied using a small antenna, and static
magnetic fields from a Ioffe--Pritchard (IP) magnetic trap. Bose
condensed $^{87}$Rb atoms are created in the $F=1$, $m_F = -1$
state in a cigar-shaped axisymmetric IP trap and then transferred
into three types of adiabatic potentials: a finite-depth
cigar-shaped harmonic trap, a cigar-shaped trap which is strongly
anharmonic, and a curved-sheet potential. The primary results
presented in this paper are measurements of the population in each
$m_F$ state for the finite depth and anharmonic potentials and
measurements of the displacement of the curved-sheet potential for
different frequencies of the RF magnetic field.

\section{Theory}

An atom moving in the presence of static and RF magnetic fields
experiences adiabatic potentials as long as there are no
Landau-Zener transitions.  These adiabatic potentials are provided
by a spatially inhomogeneous static magnetic field which gives
rise to a spatially varying detuning for the RF magnetic field. In
the adiabatic potential, the atom exists in a superposition of
hyperfine or Zeeman states, which is a stationary state in the
rotating frame. These superposition states are typically referred
to as dressed states \cite{CohenTannoudji68}.

In our experiment, the dressed states are composed of the $m_F =
\pm 1$ and $m_F = 0$ Zeeman states in the $5S_{1/2}$, $F=1$
hyperfine ground state of $^{87}$Rb.  The Zeeman states are
coupled by an RF magnetic field $\vect{B}_\textrm{rf} = \cos
(\omega t) \unitvect{y}$, where $\unitvect{z}$ is the weak
direction of the IP trap and the gravitational force acts in the
$-\unitvect{y}$ direction. The frequency $\omega$ of the RF field
is tuned near the $m_F = 0 \ToFrom m_F = +1$ and $m_F = 0 \ToFrom
m_F = -1$ transitions, which are nearly degenerate at the static
field magnitudes used here ($< 11$~G).

To determine the dressed state wavefunctions, one must solve three
coupled Gross-Pitaevskii equations. A simplification of these
equations is possible in our system because the interaction
strengths between different $m_F$ states are nearly equal, and
spin exchange rates are small compared to the RF Rabi rate (we
estimate spin-exchange rates less than 10 Hz). The many-body
wavefunction may therefore be written as $\ket{\psi} = c_+ \ket{+}
\psi_+ + c_0 \ket{0} \psi_0 + c_- \ket{-} \psi_-$, where $\ket{+},
\ket{0}, \ket{-}$ are the single atom dressed states and $\psi_+,
\psi_0, \psi_-$ are the Thomas--Fermi wavefunctions determined by
the corresponding adiabatic potentials. In the dressed-state
basis, $\ket{\psi}$ is a stationary state---the numbers of atoms
in each dressed state are independent and are determined by the
process used to transfer atoms into the adiabatic potentials.

The dressed states can be written in terms of a single angle
defined by $\tan \theta = \sqrt{2} \Omega/\delta$,
\begin{subequations} \label{eq:DressedStates} \begin{eqnarray}
 \ket{+} &=& \frac{1}{2} \left[ (1+\cos \theta)\ket{m_F = 1}
 + \sqrt{2} \sin \theta \ket{m_F = 0} \right. \nonumber \\ & & \left.
 +  \frac{\sin^2 \theta}{1+\cos \theta} \ket{m_F = -1} \right], \\
 \ket{-} &=& \frac{1}{2} \left[ (1-\cos \theta)\ket{m_F = 1}
 - \sqrt{2} \sin \theta \ket{m_F = 0} \right. \nonumber \\ & & \left.
 + \frac{\sin^2 \theta}{1-\cos \theta} \ket{m_F = -1} \right], \\
  \ket{0} &=& \frac{1}{2} \left[ - \sqrt{2} \sin \theta \ket{m_F = 1}
 + 2 \cos \theta \ket{m_F = 0} \right. \nonumber \\ & & \left.
 + \sqrt{2} \sin \theta \ket{m_F = -1} \right].
 \end{eqnarray} \end{subequations}
Here, the detuning $\delta = \omega - \omega_B$ is measured from
the frequency $\omega_B = \mu B/\hbar$ of the $\ket{m_F = \pm 1}
\ToFrom \ket{m_F = 0}$ transitions, which is determined by the
local magnetic field $B$ experienced by an atom and the atomic
magnetic moment $\mu = g_F \mu_B$ ($g_F=-1/2$ is the $g$-factor
for the $F=1$ state, and $\mu_B$ is the Bohr magneton). The Rabi
rates, $\Omega_\pm = -\frac{1}{2}\matrixel{m_F = \pm 1}{\vect{\mu}
\cdot \vect{B}_\textrm{rf}}{m_F = 0}$, for the $\ket{m_F = \pm 1}
\ToFrom \ket{m_F = 0}$ transitions depend on the static magnetic
field.  In our experiment, the rates are nearly equal: $\Omega_\pm
\approx \Omega \approx 0.483 \times \mu_B B_{\textrm{rf}}/\hbar$.
The dressed state wavefunctions are written in
Eq.~(\ref{eq:DressedStates}) in a rotating frame defined by $c_+ =
\tilde{c}_+ e^{i \omega t}$, $c_- = \tilde{c}_- e^{-i \omega t}$,
and $c_0 = \tilde{c}_0$, where the single-atom spin wavefunction
is written in the Schrodinger picture as $\ket{\tilde{\psi}} =
\tilde{c}_+ \ket{m_F = +1} + \tilde{c}_- \ket{m_F = 0} +
\tilde{c}_- \ket{m_F = -1}$. The dressed states have eigenenergies
$E_+ = \hbar \sqrt{\delta^2 + 2 \Omega^2}$, $E_- = -\hbar
\sqrt{\delta^2 + 2 \Omega^2}$, and $E_0 = 0$.

\begin{figure}
\includegraphics[scale=0.8]{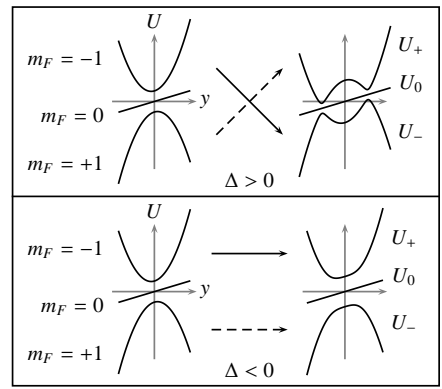}
\caption{Adiabatic potentials for the $F=1$ state in the presence
of gravity. The potential $U$ is shown along the $\unitvect{y}$
direction for the $\ket{+}$, $\ket{0}$, and $\ket{-}$ states on
the right and with no RF magnetic field present on the left.  The
arrows indicate how the $m_F$ states transform when the RF field
is turned on adiabatically for positive ($\Delta>0$) and negative
($\Delta<0$) detuning.} \label{fig:Potentials}
\end{figure}

The shape of the adiabatic potentials is determined by the effect
of the inhomogeneous magnetic field created by the IP trap, where
\begin{equation} \label{eq:IPfieldsq}
 B(\vect{r})^2 \approx B_0^2 + (C_1^2 - B_0 B_2) r^2 + 2 B_0 B_2 z^2,
\end{equation}
near the center of the trap \cite{Metcalf87}. Here, $B_0$ is the
bias field, i.e., the magnitude of the field at the center of
trap, $C_1$ is the radial gradient, and $B_2$ is the axial
curvature.  For the work presented in this paper, $B_0 = 1.33(1)
\textrm{ MHz} \times h/\mu$, and $C_1 = 355(2)$~G/cm and $B_2 =
129(2) \textrm{ G/cm}^2$ (determined from measurements of the
harmonic trapping frequencies). The adiabatic potentials for the
$\ket{+}$ and $\ket{-}$ states are $U_{\pm} = \pm \sqrt{[\mu
B(\vect{r})]^2 + 2(\hbar \Omega)^2} + mgy$, where $g$ is the local
acceleration due to gravity. The $\ket{0}$ state is only affected
by gravity and experiences a potential $U_0 = mgy$. The adiabatic
potentials are shown in Fig.~\ref{fig:Potentials}.

The shape of the adiabatic potentials depends on the detuning,
$\Delta = \omega - \omega_0$, relative to the resonance at the
trap center, $\omega_0 = \mu B_0 / \hbar$.  We summarize here the
shape of the adiabatic potentials, starting with the \Uminus\
potential. The \Uminus, $\Delta> 0$ potential is the familiar
potential utilized for forced RF-induced evaporative cooling
\cite{evapcooling}.  This potential is harmonic near the central
minimum and the curvature there is independent of detuning, except
for small (compared with the Rabi rate) detunings. The \Uminus,
$\Delta < 0$ potential is not confining.

In contrast to the \Uminus\ potential, the \Uplus\ potential is
always confining with a trapping geometry that depends strongly on
detuning: for large and negative $\Delta$, it is harmonic near the
potential minimum. For small and negative $\Delta$, the \Uplus\
potential is quartic in the absence of gravity; gravitational sag
results in an overall trapping potential for this case which is
strongly anharmonic and cannot be characterized as purely quartic.
For $\Delta > 0$, the \Uplus\ potential is an ellipsoidal shell in
the absence of gravity. Atoms in the $\ket{+}$ state are forced by
gravity to occupy a curved sheet in the bottom of this shell.

\section{Experiments}

The dressed states are primarily characterized by the population
in each $m_F$ state. We investigate the $m_F$ state composition of
the dressed states in the \Uminus\ and \Uplus\ potentials for
$\Delta> 0$ and $\Delta < 0$, respectively, by measuring the
population in the $m_F = 0$ and $m_F = -1$ states after the RF is
turned off suddenly. We also show that measuring the populations
in the $m_F$ basis for different detunings is a sensitive method
for determining the Rabi rate.

We begin by creating a condensate of approximately $5 \times 10^5$
$^{87}$Rb atoms in the $F=1$, $m_F=-1$ state with no observable
thermal component via RF evaporative cooling (we use an apparatus
similar to that described in \cite{Lewandowski03}). The atoms are
trapped in a Ioffe--Pritchard type magnetic trap as described
above with radial and axial trapping frequencies $\omega_r =
231(8)$~Hz, and $\omega_z = 14.5(3)$~Hz, respectively. Atoms are
imaged via a standard absorption imaging technique, using light
resonant with the $5S_{1/2}, F=2$ to $5P_{3/2}, F=3$ cycling
transition. Immediately before the probe beam is switched on,
atoms are optically pumped from the $F=1$ manifold to the $F=2$
manifold using light tuned to the $5S_{1/2}, F=1$ to $5P_{3/2},
F=2$ transition. RF signals are coupled to the atoms by an antenna
which is a three turn, 10 mm radius coil located 1.5 cm from the
atoms in the $\unitvect{y}$ direction.

To measure the $m_F$ state composition, the RF field is turned on
slowly to transfer all the atoms in the condensate from the $m_F =
-1$ state to a single dressed state, and then turned off suddenly
to project the dressed state on to the $m_F$ basis.  Once a
condensate is created, the RF power is increased linearly from 0
to 35~dBm (of which a fraction is coupled into the RF antenna)
over 5~ms at 1.10~MHz (1.75~MHz) for data taken with $\Delta < 0$
($\Delta > 0$). The RF frequency is swept linearly at 10~kHz/ms to
the value at which we wish to measure the spin composition; the RF
is then shut off suddenly (the RF source is turned off within
3~ns). Turning off the RF much faster than the effective Rabi rate
projects the dressed states into the $m_F$ basis, making it
possible to investigate the spin composition by measuring the
relative number of atoms in each $m_F$ state. After waiting 3~ms
for the $m_F = 0$ atoms to fall and spatially separate from the
$m_F = -1$ atoms, the magnetic trap is shut off. This 3~ms delay
also forces any $m_F = +1$ atoms to be ejected from the trapping
region by the inhomogeneous magnetic field from the IP trap. The
$m_F = 0$ and $m_F = -1$ clouds are imaged 7 ms after the magnetic
trap is turned off. We verify that the frequency sweep is
adiabatic by sweeping the RF source back to the initial frequency
(1.10~MHz or 1.75~MHz) before turning off the RF field and
observing no loss of atoms.

\begin{figure}
\includegraphics[scale=0.8]{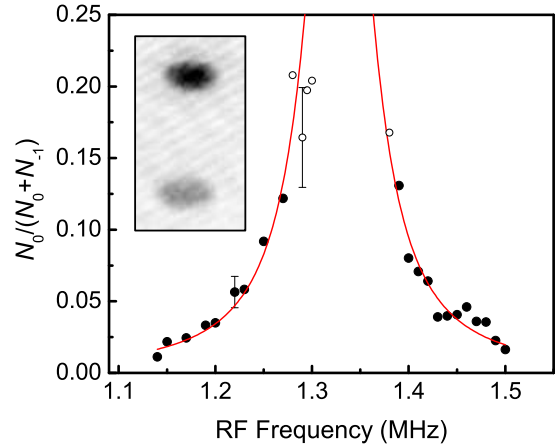}
\caption{Ratio of $m_F=0$ atoms to total number of $m_F = 0$ and
$m_F = -1$ atoms versus RF frequency. Data for RF frequencies
greater (less) than 1.328~MHz represent atoms confined in the
\Uminus\ (\Uplus) potential.  Each data point represents a single
image, except the points with error bars, which are each an
average of 7 measurements; the error bars indicate the standard
deviation of these measurements. The solid red line is a fit to
Eq.~(\ref{eq:SpinComp}). Open circles represent data points for
which gravitational sag is calculated to shift the measured ratio
by more than 1\% and are not included in the fit. The inset shows
a typical image used to obtain the data; the field of view is
approximately 260~$\mu$m by 370~$\mu$m. The lower (upper) cloud
consists of $m_F = 0$ ($m_F = -1$) atoms.}
\label{fig:SpinComp}
\end{figure}

The number of atoms in each $m_F$ state is obtained by fitting the
absorption images to the Thomas--Fermi profile expected for a BEC.
The data in Fig.~\ref{fig:SpinComp} show the experimentally
determined ratio of $m_F = 0$ atoms to the total number of $m_F =
0$ and $m_F = -1$ atoms for different final RF frequencies; as
many as 20\% of the atoms are observed in the $m_F = 0$ state. The
data in Fig.~\ref{fig:SpinComp} are fitted to
\begin{equation} \label{eq:SpinComp}
 \frac{N_{0}}{N_{0} + N_{-1}} = \frac{4 \Omega^2}{4 \Omega^2
 + (|\delta| + \sqrt{\delta^2 + 2 \Omega^2})^2},
\end{equation}
which is the relative Zeeman state distribution expected from
Eq.~(\ref{eq:DressedStates}) \cite{endnote3}.  Only points at
detunings beyond $\pm50$~kHz, where gravitational sag is
calculated to shift the measured ratio by less than 1\%, are
included in the fit. The Rabi rate and $\omega_0$ are left as free
parameters in the fit; the values obtained are $\Omega = 2 \pi
\times 24.6(5)$~kHz and $\omega_0 = 2 \pi \times 1.328(2)$~MHz,
corresponding to an RF field strength $B_{\textrm{rf}} =
36.3(7)$~mG. Determining the Rabi rate is normally very difficult
in cold-atom experiments involving transitions between
magnetically trapped and untrapped states, because Rabi
oscillations cannot be directly observed and uncertainty in the
geometry of the RF antenna. We are able to determine $\Omega$ to
within 2\% via spin composition measurements. The response of the
RF source and antenna do not vary significantly over the frequency
range covered by the data in Fig.~\ref{fig:SpinComp}.


The simplest example of using adiabatic potentials to confine
ultra-cold atoms in a novel geometry can be realized using the
$\ket{+}$ state for positive detuning. Atoms in the \Uplus\
potential for $\Delta > 0$ occupy a curved sheet, which is the
bottom part of an ellipsoidal shell potential \cite{Zobay01}. We
characterized the shape of the \Uplus\ potential by measuring the
displacement of atoms in the $\ket{+}$ state from the minimum of
the $m_F = -1$ potential (in the absence of the RF field).

\begin{figure}
\includegraphics[scale=0.9]{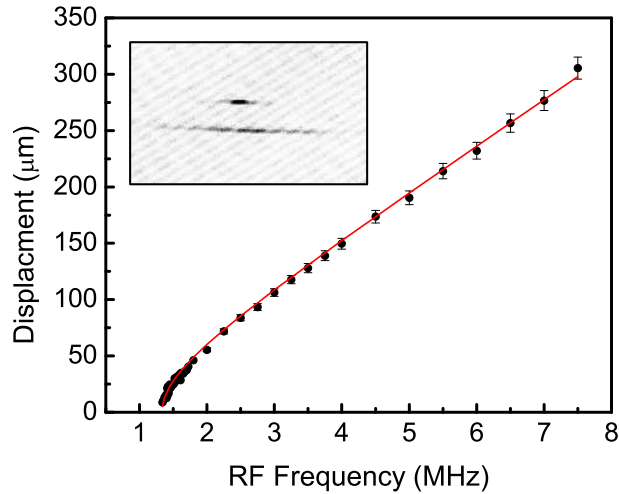}
\caption{Displacement from the IP trap center for atoms trapped in
\Uplus\ potential for different RF frequencies. The solid red line
shows the calculated displacement.  The inset shows atoms
simultaneously loaded into the \Uminus\ (upper cloud) and \Uplus\
potentials at $\Delta = 2 \pi \times 1$~MHz. The field of view for
the inset (taken while the atoms were trapped) is approximately
640~$\mu$m by 390~$\mu$m.} \label{fig:CenterDisp}
\end{figure}

Atoms are transferred into the \Uplus\ potential by turning on the
RF field at $\Delta < 0$ and sweeping the frequency through
$\omega_0$ to the value where the displacement will be measured.
\cite{Zobay01,Lorent04}. In general, care must be taken to sweep
slowly enough to adiabatically transfer a BEC into the \Uplus\
potential. We made no attempt to sweep adiabatically because the
measured displacement was found to be independent of the details
of the frequency sweep. For the data shown in
Fig.~\ref{fig:CenterDisp}, the RF is turned on at 1.10~MHz and
swept at 31~kHz/ms using $\Omega \approx 2 \pi \times 40$~kHz.
After transferring atoms into the \Uplus\ potential, the atoms are
held for 4~ms and then imaged in the trap. The center of the atom
cloud is determined by a Gaussian fit to the absorption image.
Fig.~\ref{fig:CenterDisp} shows the measured displacement from the
the minimum of the $m_F = -1$ potential for different RF
frequencies.  The error bars account for uncertainty in the
magnification of our imaging system.

The measured data agree well with the displacement expected for an
IP magnetic trap. The solid red line in Fig.~\ref{fig:CenterDisp}
is the displacement calculated using the magnetic field minimum in
Eq.~(\ref{eq:IPfieldsq}) at $z = 0$. We have ignored gravitational
sag, calculated to be less than 1~$\mu$m for detunings beyond
0.2~MHz. At large detunings, the displacement is dominated by the
radial gradient of the IP trap.  For small detunings, the cloud
samples only the field near the center of the IP potential, and
the displacement is quadratic in detuning. Plotting
Fig.~\ref{fig:CenterDisp} as RF frequency vs.~displacement
produces a map of the radial field of the IP trap. This provides a
method to measure the radial field to high accuracy---fitting the
data in Fig.~\ref{fig:CenterDisp} to the equation used to
calculate the solid red line, but leaving $\omega_0$ and $C_1$ as
free parameters, yields $C_1 = 357(2)$~G/cm.  The radial gradient
obtained using this technique is in excellent agreement with the
value derived from measuring the IP trap frequencies through
``sloshing" motion of the condensate.

We have also realized an unusual trapping geometry in which a
cigar-shaped cloud of atoms is trapped above a curved sheet of
atoms. By increasing the frequency sweep rate and/or decreasing
the RF power, it is possible to transfer only a fraction of the
atoms to the \Uplus\ potential. Under these conditions,
Landau-Zener transitions result in atoms trapped in both the
\Uplus\ and \Uminus\ potentials, as seen in the inset to
Fig.~\ref{fig:CenterDisp}. Here, the detuning was $\Delta = 2 \pi
\times 1$~MHz, and the loading ramp was altered so that the
frequency sweep rate was 300~kHz/ms and $\Omega = 2 \pi \times
3.3$~kHz within $\pm 40 \textrm{ kHz} \times 2 \pi$ of $\omega_0$.

We briefly discuss the prospects for obtaining a 2D BEC in the
\Uplus, $\Delta > 0$ potential. Motion around the bottom of this
potential is characterized by three different harmonic oscillator
frequencies: a large radial frequency and two small frequencies
along directions which experience pendulum-like motion
\cite{Zobay04,Lorent04}. For the \Uplus\ potential at $\Delta = 2
\pi \times 1$~MHz and $\Omega = 2 \pi \times 40$~kHz, the radial
and two pendular frequencies are calculated to be 900~Hz and 60~Hz
and 4~Hz, respectively.  These frequencies are calculated using a
classical normal-modes calculation and the magnetic field from
Eq.~(\ref{eq:IPfieldsq}). We observed a lifetime of 4(2)~s and a
heating rate of 4.5(3)~$\mu$K/s under these conditions. Sources of
atom loss may be Landau--Zener transitions to untrapped states
\cite{Zobay04}; three body recombination may also play a role
because the non-adiabatic loading into this potential can
transiently increase the density \cite{endnote2}. Because
adiabatic transfer into the the 2D regime is quite slow, due to
the low pendular trap frequencies of the \Uplus\ potential, we
expect obtaining a condensate in the 2D regime to be difficult
without significant improvement to the heating rate. A large
heating rate was also observed in \cite{Lorent04}, where it was
attributed to noise in the RF frequency.  Finally, we remark that
we have observed partially condensed clouds upon release from the
\Uplus\ potential for $\Delta$ up to 1.5~MHz after rapid
(non-adiabatic) loading (31~kHz/ms frequency sweep rate),
indicating that the condensate is not destroyed immediately by the
loading process.



\section{Conclusion}

In conclusion, we have investigated the nature of RF dressed
potentials for magnetically trapped $^{87}$Rb in the $F=1$ state
and measured the $m_F$ state composition of the dressed states. We
have also studied the shape of the ellipsoidal \Uplus\ potential.
Potentially useful techniques for measuring the RF Rabi rate and
for directly characterizing the radial field of a magnetic trap
were also demonstrated.

Future directions for this research include creating ring traps
\cite{Garraway05,Schmiedmayer06}, which are promising for use as
atomic Sagnac interferometers. Courteille et.\ al.\
\cite{Zimmermann06} also recently proposed a scheme to create
versatile potentials using multiple RF frequencies. An important
future step is to characterize and address heating and loss
mechanisms in dressed state potentials. Because the dressed states
are superpositions of the atomic spin states, a related question
of particular interest concerns the nature of spin exchange
collisions for dressed states.

We acknowledge funding from the Office of Naval Research (Grant
No. N000140410490), the National Science Foundation (Award No.
0448354), the Sloan Foundation, and the University of Illinois
Research Board.  We thank Deborah Jin for a careful reading of
this manuscript.




\begin{thebibliography}{99}
\bibitem{endnote1} The combined RF-frequency and static magnetic
potential is smooth provided that the trapping region is much
smaller than the characteristic dimensions of the magnetic trap.
\bibitem{endnote3} We plot the ratio $N_0/(N_0+N_{-1})$ because we cannot measure the
number of atoms in the $m_F=1$ state, and hence we cannot
determine the total number of atoms.
\bibitem{endnote2} Based on the calculated trap frequencies, the equilibrium density is expected
to be lower in the adiabatic potential than in the IP trap.
\end{thebibliography}
\end{document}